# Observation of the Fractional Quantum Hall Effect in Graphene

Kirill I. Bolotin*[1], Fereshte Ghahari*[1], Michael D. Shulman[2], Horst L. Stormer[1, 2] & Philip Kim[1, 2]

[1]Department of Physics, Columbia University, New York, New York 10027, USA  [2]Department of Applied Physics and Applied Mathematics, Columbia University, New York, New York 10027, USA

*These authors contributed equally to this work.

**When electrons are confined in two dimensions (2D) and subjected to strong magnetic fields, the Coulomb interactions between them become dominant and can lead to novel states of matter such as fractional quantum Hall (FQH) liquids[1]. In these liquids electrons linked to magnetic flux quanta form complex composite quasipartices, which are manifested in the quantization of the Hall conductivity as rational fractions of the conductance quantum. The recent experimental discovery of an anomalous integer quantum Hall effect in graphene has opened up a new avenue in the study of correlated 2D electronic systems, in which the interacting electron wavefunctions are those of massless chiral fermions[2,3]. However, due to the prevailing disorder, graphene has thus far exhibited only weak signatures of correlated electron phenomena[4,5], despite concerted experimental efforts and intense theoretical interest[6-12]. Here, we report the observation of the fractional quantum Hall effect in ultraclean suspended graphene, supporting the existence of strongly correlated electron states in the presence of a magnetic field. In addition, at low carrier density graphene becomes an insulator with an energy gap tunable by magnetic field. These newly discovered quantum states offer the opportunity to study a new state of matter of strongly correlated Dirac fermions in the presence of large magnetic fields.**

In a perpendicular magnetic field, the energy spectrum of a clean two-dimensional electron system (2DES) splits into a fan of Landau levels (LLs). When the Fermi energy is tuned to lie between the LLs, the system enters the integer quantum Hall (IQH) regime, in which current is carried by states at the edge of the sample and the overall conductance is quantized as $G=ve^2/h$, where $v$ is an integer Landau level filling factor. Already the first observation of the IQH effect in graphene demonstrated an unusual sequence of filling factors $v=\pm2,\pm6,\pm10...$, which differs from previously studied 2DESs. This sequence originates from two peculiar features of graphene: the four-fold spin and pseudo-spin (valley) degeneracy of LLs and the existence of a non-trivial Berry phase associated with the pseudo-spin of Dirac quasiparticles[13].

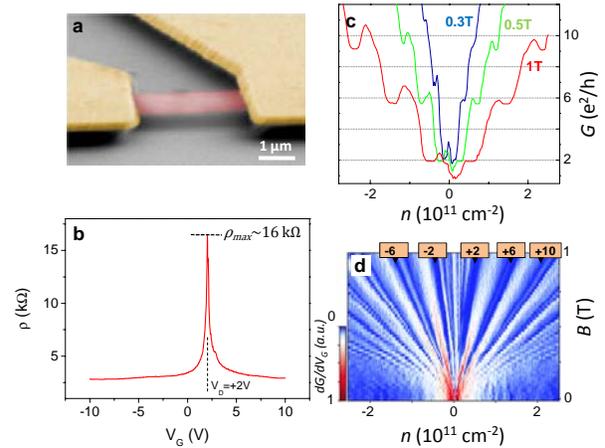

**Figure 1 Electrical properties of suspended graphene at low magnetic field a,** False-colour scanning electron micrograph of a typical device. Single layer graphene (red) is suspended ~150 nm above $SiO_2$/Si substrate (gray) and supported by two gold electrodes (yellow). To ensure mechanical stability of the devices we use a two-terminal configuration in this study. **b,** Resistivity $\rho$ of graphene as a function of gate voltage $V_G$ applied between the graphene and the substrate is measured at $T=2$ K and $B=0$ T. The resistivity is calculated from the resistance $R$ via $\rho=R\times W/L$, where $W$ is sample width and $L$ – length. **c,** Conductance $G$ as a function of carrier density $n$ at different magnetic fields $B=0.3, 0.5, 1$ T. **c,** Colour rendition of abs[$dG/dn(B,n)$] in the low magnetic field regime $B=0$-1 T. Dark colour corresponds to quantum Hall plateau region $G(n)=ve^2/h=$const. The gate capacitance $C_g$~50 aF/$\mu m^2$ used in calculating the carrier density $n=C_g(V_g-V_D)$ is obtained from the shift of the QH plateaus with magnetic field.

In clean samples and under very strong magnetic fields, additional IQH states emerge at filling factors $v=0, \pm1$[4,5]. These fragile IQH states are conjectured to result from electron-electron (e-e) interactions lifting the pseudospin and spin degeneracy of the zeroth LL[14]. The nature of these states, and in particular the unusual $v=0$ state, responsible for the divergent resistivity of graphene at high magnetic fields[15], has raised considerable interest in the effect of e-e interactions among Dirac quasiparticles. In conventional semiconductor heterojunctions such correlation effects are spectacularly manifested in the FQH effect, where new electronic ground states are formed in which the elementary excitations are composite particles with fractional charge[16]. The possibility of a FQH effect in graphene, and the interplay between many-body correlations and the Dirac nature of quasiparticles has been predicted to result in rich physics[6-12]. Thus far, however, strong electron scattering off omnipresent residual impurities have precluded the observation of the FQH effect.

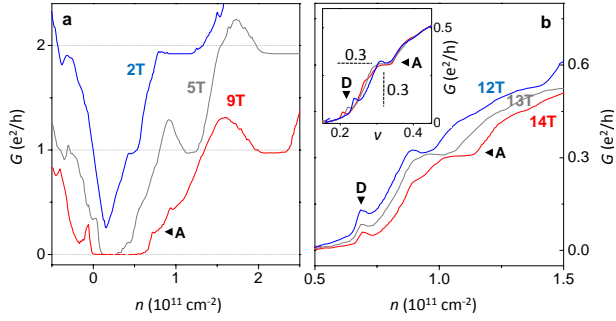

**Figure 2 Magnetotransport at high magnetic fields a,** $G(n)$ at $B$=2, 5, 9 T. QH state $v$=1 QH develops at $B$=2 T, features with $v$<1 form at 9 T. **b,** $G(n)$ for $B$=12, 13, 14 T. The data are acquired at $T$=6 K to suppress resistance fluctuations associated with quantum interference. A plateau 'A' with $G$~0.3 $e^2/h$ and at a density dependent on $B$ emerges when $B$>11 T. Another feature, 'B' appears at constant $n$ at different fields. Inset, $G$ is replotted as a function of filling factor $v=n\varphi_0/B$ for $B$=12, 13, 14 T. At different fields, the $G(v)$ curves around 'A' collapse onto a single universal trace allowing identification of feature 'A' as the FQH state $v$=1/3.

It has been previously demonstrated[17,18] that the scattering can be drastically reduced by fabricating suspended graphene devices. In such devices, the carrier mobility $\mu=(ne\rho)^{-1}$, where $n$ is the carrier density and $\rho$ is the resistivity, can exceed 200,000 cm$^2$/Vs. This suggests that transport in the extreme quantum limit could be probed for the FQH effect. In this study we use two-terminal suspended graphene devices (Fig. 1a. See the supplementary information for fabrication details). These devices mechanically more robust and tend to better survive thermal cycling, compared to the previously studied multiprobe specimens[17]. The devices are measured in a cryostat capable of magnetic fields $B$ up to 14 T and temperatures $T$ between 2 and 15 K. The two-terminal conductance $G$ is recorded as a function of $B$, $T$, and the carrier density $n$, which is tuned by adjusting the voltage $V_G$ between the graphene and the back gate electrode. We limit $|V_G|$<10 V in order to avoid collapse of the graphene due to electrostatic attraction, limiting the accessible carrier densities to $|n|$ < $3.5\times10^{11}$ cm$^{-2}$. At $T$=2 K a sharp peak in resistance $R=1/G$ is evident around the Dirac point $V_D \sim 2$ V; at this gate voltage graphene is charge neutral (Fig. 1b). From the observed full width at half maximum of $R(V_g)$, we estimate the density inhomogeneity of the suspended graphene to be less than $1\times10^{10}$ cm$^{-2}$ [17].

We first confirm the high quality of the suspended graphene devices by studying them at low magnetic fields. Remarkably, even at $B$=0.3 T, we observe a developing plateau in $G$ at $2e^2/h$ associated with the IQH effect at $v$=2. At B=1T, fully developed plateaus are evident at $v$=±2, ±6, ±10 (Fig. 1c). It is convenient to visualize the IQH effect by means of a Landau fan diagram, where the derivative $|dG/dn(B,n)|$ is plotted as a function of $B$ and $n$ (Fig 1c). At a QH plateau corresponding to filling factor $v$, the conductance remains constant $G(V_G)=ve^2/h$, and the electron density is given by $n(B)=Bv/\varphi_0$, where $\varphi_0=h/e$ is the magnetic flux quantum. Therefore, QH plateaus $dG/dn=0$ appear as stripes fanning out from point B=0, $V_G=V_D$ and with slope $dn/dB=v/\varphi_0$. We observe such stripes down to $B$<0.1 T, indicating that the QH effect survives to very low fields. Assuming $\mu B >>1$ for the IQHE to exist[17], this observation implies a lower bound for the mobility of ~100,000 cm$^2$/Vs.

In such ultraclean graphene samples the extreme quantum limit can be reached at relatively low magnetic field. In the intermediate field regime $B$<10 T (Fig. 2a), we observe an IQH plateau corresponding to $v$=1 at a field as low as $B$=2 T, in addition to the well-resolved $v$=2 IQH plateau already mentioned. Similarly, a $G$=0 plateau appears around the region of zero density ($V_G=V_D$) for $B$>5 T, indicating the onset of an insulating state at low density. In the multi-terminal Hall bar samples, this insulating behaviour is considered a signature of the $v$=0 IQH effect since it accompanies a plateau of Hall conductivity $\sigma_{xy}$=0. In samples on a substrate[4,5], such $v$=0, ±1 IQH plateaus could only be observed for $B$>20 T. These plateaus are believed to arise from a magnetic field induced spontaneous symmetry breaking mediated by e-e interactions[14]. This suggests that e-e correlations in ultraclean suspended graphene can be sustained at much lower fields as compared to graphene on substrates. The strength of e-e interactions in a filled LL can be estimated to be proportional to $e^2/l_b$ [16], where $l_b=(\hbar/eB)^{1/2}$ is the magnetic length, and thus increases with $B$. We therefore expect stronger interaction effects to occur at higher magnetic fields.

Indeed, two notable features emerge in the high magnetic field regime, B>10T: a peak-like and a plateau-like structure marked by 'D' and 'A' respectively (Fig. 2a). While the position of the peak-like feature 'D' remains unchanged, feature 'A' moves to higher densities with increasing $B$, developing into a well-defined plateau. Since the density $n$ corresponding to 'D' does not change with $B$, this feature is likely due to $B$-independent rapid threshold fluctuations, associated with hopping or resonant tunnelling through localized states[19]. On the other hand, the electron density corresponding to the plateau-like structure 'A' changes with increasing $B$, suggesting that 'A' is related to a QH state with $v$>0 [20].

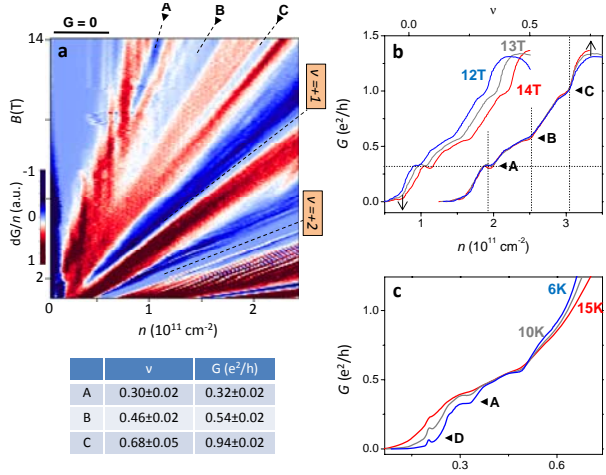

**Figure 3 Identifying additional FQH states a,** Landau fan diagram $|dG/dn(B,n)|$ at $T=6$ K. Dark blue colour corresponds to $G(n)=\nu e^2/h=$const. At high fields features 'B' and 'C' with $\nu=0.46$, 0.68 respectively emerge in addition to feature 'A' (FQH $\nu=1/3$). Insulating $G=0$ state is evident around $n=0$ at $B>5$ T. The data are acquired from the same sample as Fig. 2, but in a different thermal cycle, where disorder related 'D' is absent. For clarity, only the electron part of the data set $n>0$ is shown, the data for the hole side is shown in the Supplementary Information. **b,** $G$ as a function of both $n$ (bottom axis) and filling factor $\nu$ (top axis) at $B=12$, 13, 14 T. Features 'A', 'B', and 'C' in curves at different fields collapse onto a single universal curve $G(\nu)$. **c,** $G(\nu)$ at different temperatures $T=6$, 10, 15 K. Feature 'A' (FQH state $\nu=1/3$) survives up to $T=10$ K. **Inset,** Filling factors $\nu$ and conductance $G$ values corresponding to features 'A','B', and 'C'.

A consistent picture emerges when we plot $G$ as a function of filling factor $\nu=n\,\varphi_0/B$ (Fig 2b, inset). The traces $G(\nu)$ acquired at different magnetic fields collapse into a single universal curve around the feature 'A'. The associated plateau has a value of $G=0.32\pm0.02$ $e^2/h$ and is positioned around $\nu=0.30\pm0.02$. In contrast, the region around 'D' does not collapse onto a universal curve and is therefore of different origin. We assign the observed plateau 'A' to the hitherto unobserved[6-12] FQH state at $\nu=1/3$. We note that at our highest field, $B=14$T, we observe the $\nu=1/3$ plateau at $n\sim10^{11}$ cm$^{-2}$, more than an order of magnitude higher than the estimated density inhomogeneity in our device. This establishes that our observation of fractional quantization is not caused by the addition of quantum resistance observed in graphene p-n-p junctions[21,22].

Fig. 3a shows our analysis of the entire data set in terms of a Landau fan diagram over the whole experimentally accessible range of $0<B<14$ T and $|n|<3.5\times10^{11}$ cm$^{-2}$. While features stemming from localized states change between thermal cycles, the familiar IQH states at $\nu=\pm1,\pm2,\pm6$, appear consistently as the blue fan stripes with slope $dB/dn=\varphi_0/\nu$. In addition to the $\nu=1/3$ FQH state 'A', we find features fanning out with distinctly different slopes corresponding to $\nu<1$. These features, marked 'B' and 'C' appear only at low density and at high fields, yielding a kink-like structures in $G(n)$ (Fig3b). We extract $\nu=0.46\pm0.02$, and $\nu=0.68\pm0.05$ for the state B and C respectively from the slopes of the corresponding lines in Fig. 3a. As before, $G(n)$ collapses onto a universal curve around 'B' and 'C' at different fields allowing the extraction of $G=0.54\pm0.02$ $e^2/h$ and $0.94\pm0.02$ $e^2/h$, where the value of conductance is taken at the centre of each feature.

These features suggest additional FQH states, despite being less developed than the plateau at $\nu=1/3$ ('A'). Since the conductance values of each features are 20-40% larger than the expected $\nu e^2/h$ with $\nu=0.46$ ('B') and 0.68 ('C'), it is difficult to assign them to particular FQH states with certainty. We note however, that features 'B' and 'C' occur near $\nu=1/2$ and $\nu=2/3$. It is likely that the feature C is related to the $\nu=2/3$ FQH state, which is expected to be one of the strongest FQH states in graphene[7]. In that case, the deviation of conductance from the expected value can be ascribed to a sample specific effect of the two-terminal measurement, which can yield $G$ values higher than those of traditional 4-probe measurements[23]. The origin of feature 'B' is even more unclear. Judging from FQHEs in traditional 2DESs, the next state to expect is $\nu=2/5$. Yet its position around $\nu=1/2$ rather suggests it might be the two-probe signature of a broad minimum in $\rho_{xx}$ often observed in traditional 2DESs in the vicinity of $\nu=1/2$, and not a FQH state[24].

The features A, B, and C persist as the temperature is increased. While most other features associated with localized states disappear at $T\sim6$ K, the plateau corresponding to $\nu=1/3$ survives up to $T\sim10$K (Fig. 3c). This energy scale is higher than the energy scale of the FQHE observed in typical semiconductor heterojunctions with similar mobility, where the 1/3 plateau disappears for T>1 K at B=15 T[1]. This remarkable resilience of the FQHE in graphene can be ascribed to the enhanced the e-e interaction in suspended graphene due to a reduced dielectric screening. The dielectric constant $\varepsilon$ enters the characteristic energy scale of e-e interaction as $E_{e\text{-}e}=e^2/\varepsilon l_b$. In suspended graphene $\varepsilon\sim1$, as compared to $\varepsilon=13.6$ in AlGaAs/GaAs[16], leading to much stronger e-e interactions in graphene, and correspondingly, to larger energy gaps and more robust FQHEs. In fact, the theoretical expectation for the excitation energy of $\nu=1/3$ FQHE in graphene is $0.017e^2/\varepsilon l_b \sim 20$ K [7], which agrees with our observations.

Finally, we consider the insulating state that appears in the low density regime at $|\nu|<1/3$. The device is fully insulating ($R>10$ GΩ) at magnetic fields $B>5$ T and at filling factors $\nu<0.15$. In the Landau fan diagram (Fig. 3a) this insulating state is apparent as a region of zero conductance $dG/dn=0$ near the charge neutrality point. To elucidate the nature of this insulating state, we examine the temperature dependence of the device's resistance measured with a small bias voltage ($V_b<0.1$ mV). While at the highest attainable field ($B=14$T) and low temperatures ($T<10$ K) the device is fully insulating, at higher temperatures we observe a clear Arrhenius behaviour, $R(n=0.17\times10^{11}$ cm$^{-2})\sim exp(E_A/2kT)$, with an activation energy $E_A \sim 60$ K (Fig 4c, Inset). Similar analyses at different magnetic fields reveal that $E_A$ rapidly decreases with decreasing $B$ (Fig. 4c).

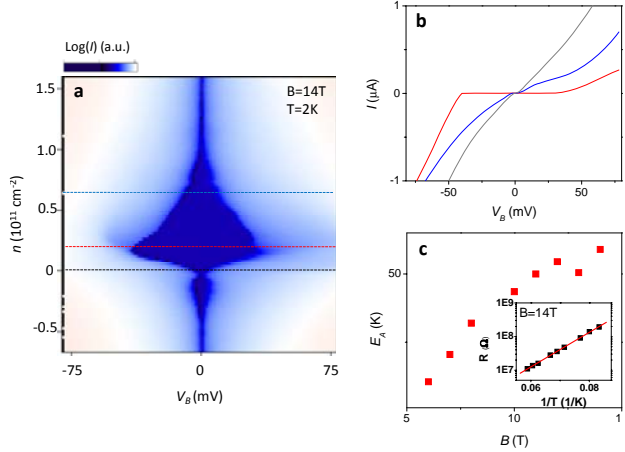

**Figure 4 The insulating state in graphene near zero density a,** Colour rendition of logarithm of the absolute value of current $I$ across the device as a function of bias voltage $V_B$ at $B=14$ T and $T=2$ K. The insulating state $I=0$ is evident as a dark blue region near $V_B=0$. **b,** $I(V_B)$ at $n=0$ (red), $0.17\times10^{11}$ cm$^{-2}$ (blue), $1\times10^{11}$ cm$^{-2}$ (black) **c,** Activation energy $E_A$ at $n=0.17\times10^{11}$ cm$^{-2}$ as a function of applied magnetic field $B$. **Inset,** Logarithm of the resistance $R$ vs. inverse temperature $1/T$ at $B=14$ T and $n=0$. The linear fit yields $R\sim exp(E_A/2kT)$ with $E_A \sim 60$ K.

Such a highly resistive state has been previously observed in more disordered graphene on substrates[15]. While its nature is hotly debated[14], the consensus is that this state stems from the symmetry breaking of the zeroth LL by e-e interactions. The activated behaviour clearly observed in our ultra-clean suspended graphene suggests the presence of a gap in the density of extended states near the Dirac point. Such a gap opening can also be probed by measuring the current $I$ as a function of the source-drain bias voltage $V_B$. Figure 4a shows $\log|I|$ as a function of both $V_B$ and $V_G$. A dark region in the central portion of this plot near $V_G=V_D$ indicates a region with near-zero conductance. Several representative $I$-$V_B$ curves at different densities (Fig. 4b) exhibit distinctly non-linear behaviour. In particular, near the charge neutrality point $V_g=V_D$, $I$ remains close to zero over a wide bias range $|V_b| < 50$ mV. This is much larger than the energy scale corresponding to the activation energy, ~4 meV, suggesting that at large bias the charge transport occurs across weakly connected insulating regions.

In conclusion, we demonstrate strongly correlated states of Dirac fermions in ultraclean suspended graphene. We observe a state at fractional filling factor $\nu=1/3$ and candidates for other possible FQHE states. At low density, graphene develops a fully insulating state whose energy gap can be tuned by magnetic field. These strongly correlated states in graphene might provide a robust basis for the implementation of exotic device schemes, such as topologically protected quantum computing utilizing FQH states[25].

**Supplementary Information** accompanies the paper on *Nature*'s website (http://www.nature.com).

**Acknowledgements** We thank Dmitry Abanin, Aron Pinczuk, and Ben Feldman for discussion. We acknowledge Andrea Young, Paul Cadden-Zimansky and Vikram Deshpande for careful reading of the manuscript. We especially thank E. Andrei for discussing her unpublished results and sample fabrication. This research was supported by the Microsoft Project Q (H.L.S) and the Department of Energy (P.K.).

**Author Contributions** K.I.B. and F.G. performed the experiments and analyzed the data. M.D.S. assisted with fabrications. H.L.S and P.K. conceived the project. All authors contributed to writing the manuscript.

**Author Information** Reprints and permissions information is available at npg.nature.com/repintsandpermissions. The authors declare no competing financial interests. Correspondence and requests for materials should be addressed P.K. (e-mail: pk2015@columbia.edu).